\ifdefined\pdfoutput
\pdfoutput=1
\fi

\documentclass[11pt]{article}

\usepackage[review]{ACL2023}

\usepackage{times}
\usepackage{latexsym}

\usepackage[T1]{fontenc}

\usepackage[utf8]{inputenc}

\usepackage{microtype}

\usepackage{inconsolata}
\usepackage{amsmath}
\usepackage{algorithm}
\usepackage{algorithmic}
\PassOptionsToPackage{table,xcdraw}{xcolor}
\usepackage{xcolor}
\usepackage{booktabs}
\usepackage{float}
\usepackage{multirow}
\usepackage{makecell}
\usepackage{graphicx}
\usepackage{amssymb} 
\usepackage{mathtools,bbm}
\usepackage{algorithm}

\usepackage{lineno}
\title{
\hspace*{-1.5cm} 
\raisebox{-0.4cm}{\includegraphics[width=1.2cm]{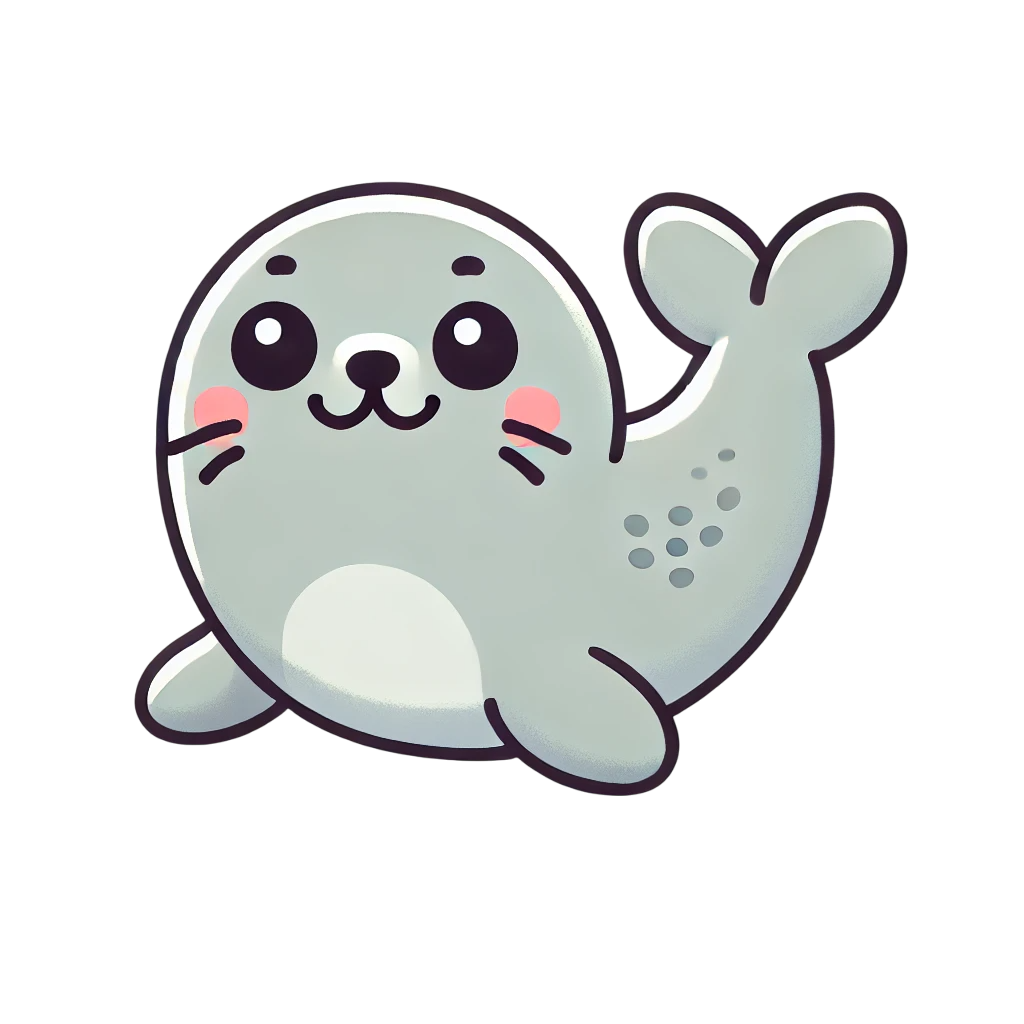}} 
SEAL: Speech Embedding Alignment Learning for Speech Large \\ Language Model with Retrieval-Augmented Generation}

\author{
  \textbf{Chunyu Sun}\textsuperscript{$1,\dagger$}, 
  \textbf{Bingyu Liu}\textsuperscript{$1,\dagger$}, 
  \textbf{Zhichao Cui}\textsuperscript{$1,\dagger$}, 
  \textbf{Junhan Shi}\textsuperscript{$1,\dagger$}, \\
  \textbf{Anbin Qi}\textsuperscript{$1$}, 
  \textbf{Tian-Hao Zhang}\textsuperscript{$1$}, 
  \textbf{Dinghao Zhou}\textsuperscript{$1$}, 
  \textbf{Lewei Lu}\textsuperscript{$1$} \\
  \textsuperscript{$1$}SenseTime Research \\
  \small\texttt{{\{sunchunyu, liubingyu, cuizhichao, qianbin, zhangtianhao1, zhoudinghao, luotto\}@sensetime.com}}
  \thanks{\,\,$\dagger$\,These authors contributed equally.}
}

\begin{document}
\maketitle
\begin{abstract}
Embedding-based retrieval models have made significant strides in retrieval-augmented generation (RAG) techniques for text and multimodal large language models (LLMs) applications. However, when it comes to speech larage language models (SLLMs), these methods are limited to a two-stage process, where automatic speech recognition (ASR) is combined with text-based retrieval. 
This sequential architecture suffers from high latency and error propagation. To address these limitations, we propose a unified embedding framework that eliminates the need for intermediate text representations. Specifically, the framework includes separate speech and text encoders, followed by a shared scaling layer that maps both modalities into a common embedding space. Our model reduces pipeline latency by 50\% while achieving higher retrieval accuracy compared to traditional two-stage methods. We also provide a theoretical analysis of the challenges inherent in end-to-end speech retrieval and introduce architectural principles for effective speech-to-document matching. Extensive experiments demonstrate the robustness of our approach across diverse acoustic conditions and speaker variations, paving the way for a new paradigm in multimodal SLLMs retrieval systems.

\end{abstract}

\section{Introduction}
In recent years, embedding models have made remarkable strides, becoming foundational components for a wide range of natural language processing tasks\citep{su2023one} \citep{cpack} \citep{wang2023improving}. These models excel at capturing semantic relationships within text data, enabling robust retrieval and matching capabilities. The success of text embeddings has naturally extended to explorations in multimodal domains, leading to significant advancements in text-image embedding models that effectively align semantic spaces across modalities \citep{chen2022murag} \citep{yasunaga2023retrieval} \citep{hu2023reveal}.
However, despite the pervasive presence of speech content in real-world applications, the integration of speech modalities into embedding-based retrieval systems remains underexplored, highlighting a critical gap in multimodal information processing.

\begin{figure}[t]
    \centering
    \includegraphics[width=1\columnwidth]{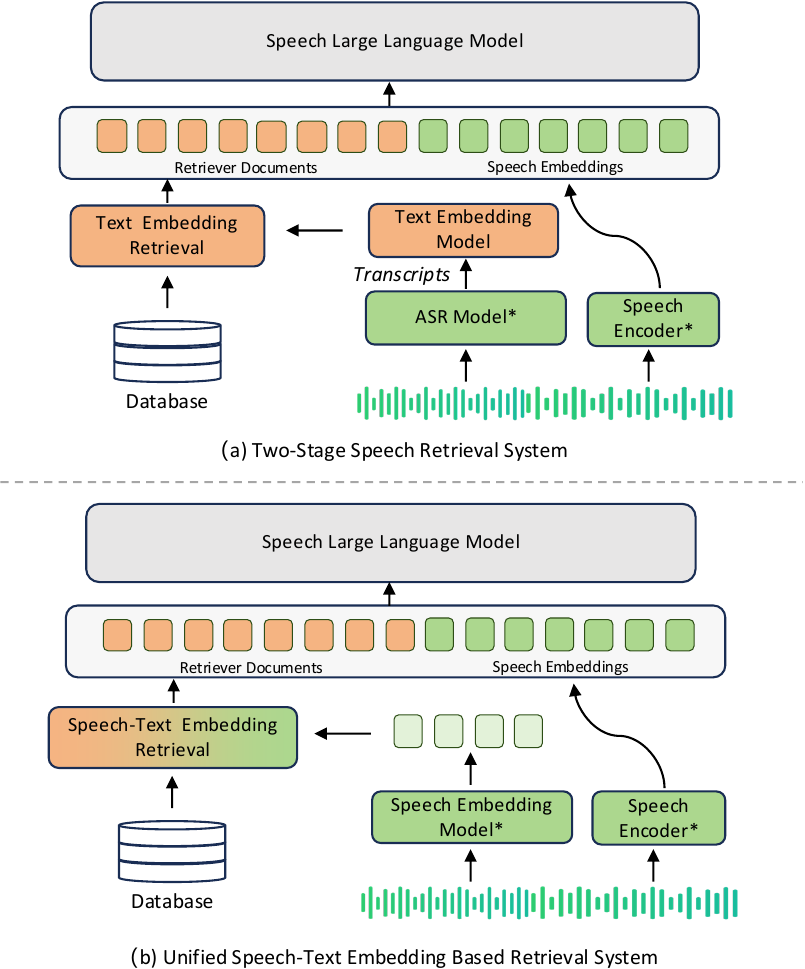}
    \caption{The traditional two-stage speech retrieval system and our proposed unified speech-text embedding based retrieval system, where * indicates that some modules in the model can be shared.}
    \label{fig:intro-figure}
\end{figure}

This challenge becomes particularly pronounced in speech-based retrieval-augmented generation (RAG) systems, which have emerged as a promising approach to augment speech large language models (SLLMs) with external knowledge \citep{jacqmin2023olisia}. Current speech RAG systems predominantly adopt a two-stage paradigm: automatic speech recognition (ASR) followed by text-based RAG as shown in Fig \ref{fig:intro-figure}. While this approach leverages well-established components, it suffers from two major limitations. First, the sequential processing of ASR and text-based RAG significantly increases system latency, rendering it impractical for real-time applications where rapid responses are critical. Second, and more fundamentally, this cascaded architecture is prone to error propagation. ASR errors—particularly in challenging acoustic environments or with non-standard speech—inevitably degrade downstream retrieval performance. This cascading effect can result in entirely irrelevant document retrieval, even when ASR errors impact only a few key terms.

Several attempts have been made to address these challenges through intermediate solutions, such as optimizing ASR for retrieval-specific metrics or implementing early fusion techniques \citep{serdyuk2018towards} \citep{szymanski2023aren}. However, these approaches still maintain the fundamental separation between speech processing and document retrieval, failing to capture the direct relationship between acoustic patterns and document semantics. Furthermore, existing methods often struggle with the inherent temporal nature of speech and the challenge of aligning variable-length speech sequences with document embeddings.

To overcome these limitations, we propose an end-to-end speech RAG model that directly learns to map speech inputs to relevant document embeddings, enhancing the performance and scalability of SLLMs. Our approach begins with separate speech and text encoders to extract features from each modality, followed by a shared scaling layer to produce the final unified speech-text embeddings. By learning a unified embedding space that captures both acoustic and semantic features, our method eliminates the need for intermediate text representations. Through careful architectural design and innovative training strategies, we achieve robust speech-to-document matching which is resilient to acoustic variations and speaker differences. Extensive experiments demonstrate that our model not only reduces pipeline latency by 50\% but also achieves higher retrieval accuracy compared to traditional two-stage systems. Our contributions open new possibilities for real-time speech-based information retrieval and challenge the conventional wisdom about the necessity of intermediate text representations in processing pipelines of RAG based SLLMs. The proposed approach not only advances the state-of-the-art in speech RAG systems but also provides a foundation for future research in end-to-end multimodal retrieval systems.

\section{Related Works}
\subsection{Multimodal Large Language Models}
Recent advances in multimodal large language models (MLLMs) \citep{li2022blip} \citep{li2023blip} \citep{liu2024visual} have demonstrated impressive capabilities in processing visual and linguistic information. Models like LLaVA \citep{liu2024visual} have established effective architectures combining LLMs with visual encoders through carefully designed projection layers. This architecture typically involves a two-stage training process: first aligning visual-textual representations, then fine-tuning on instruction data. While most efforts have focused on vision-language models, some recent work has explored speech-language models by adapting similar architectures \citep{zhang2023speechgpt} \citep{fang2024llama}. However, these models primarily focus on speech understanding and generation rather than creating robust speech embeddings for retrieval tasks. Our work bridges this gap by introducing an embedding-focused architecture specifically designed for speech-text alignment.

\subsection{Multimodal Embeddings}
In the field of multimodal embeddings, CLIP \citep{radford2021learning} has established a strong foundation for vision-language understanding through contrastive learning on large-scale image-text pairs. Its success has inspired numerous follow-up works in visual-textual alignment and retrieval. While vision-language embedding models have made significant progress, the speech domain remains relatively unexplored. Current speech-text retrieval systems typically rely on a cascaded approach, first converting speech to transcripts through ASR \cite{Hinton2012, Prabhavalkar2023, DBLP:conf/icassp/ZhangZZZL24} before applying text-based retrieval methods. This two-stage process, while straightforward, introduces latency and error propagation issues. The lack of end-to-end speech embedding models, analogous to CLIP's role in vision, represents a significant gap in multimodal understanding. Our work addresses this limitation by introducing a unified speech embedding framework that directly aligns speech with textual content, demonstrating that end-to-end speech-text embeddings can achieve superior performance while reducing computational overhead.

\begin{figure*}[t]
    \centering
    \includegraphics[width=2\columnwidth]{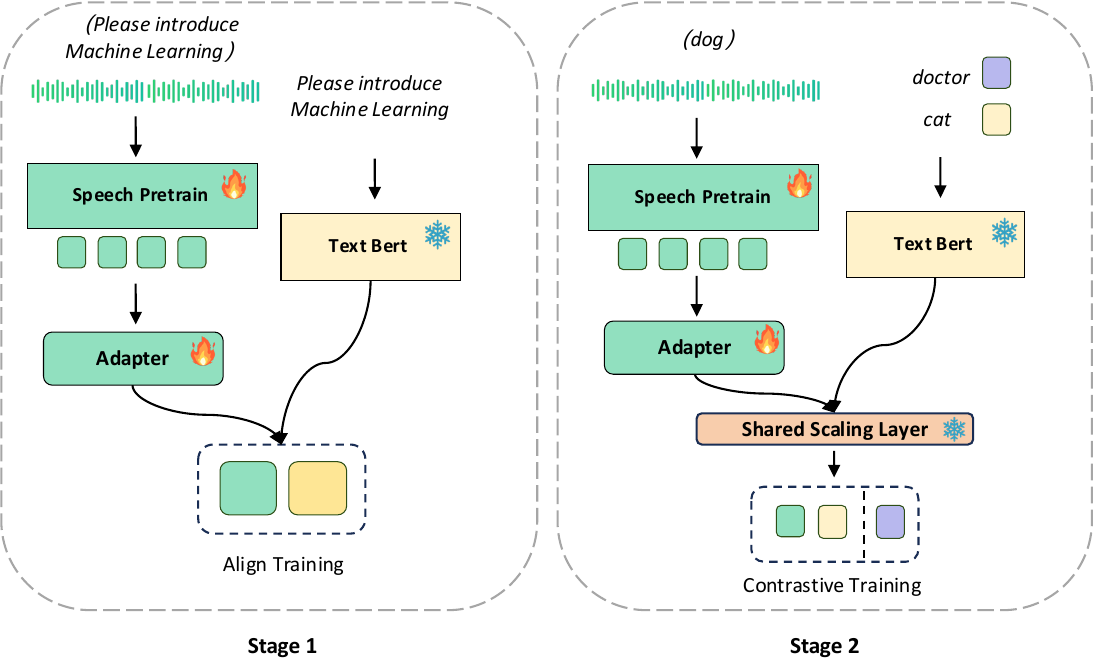}
    \caption{Overview of Our methods. The stage 1 is align training and the stage 2 is contrastive training}
    \label{fig:main-figure}
\end{figure*}

\section{Method}

In this section, we introduce our proposed method (Fig \ref{fig:main-figure}) and the two baseline methods we experimented with.
\subsection{Training Workflow}

\subsubsection{Speech-Text Alignment Pre-Training}
We propose a two-stage training approach, with pre-training focusing on aligning speech and text embeddings in a unified semantic space. Given an input pair of speech signal $x_a$ and its corresponding text transcript $x_t$, we first extract their respective features:
\begin{equation}
h_s = f_s(x_s) \in \mathbb{R}^{T \times d_s}
\end{equation}
\begin{equation}
z_t = f_t(x_t) \in \mathbb{R}^{L \times d_t}
\end{equation}
where $f_s$ and $f_t$ denote the speech and text encoders. $T$ and $L$ denote the sequence lengths of speech and text features respectively, while $d_s$ and $d_t$ represent their feature dimensions.

To bridge the modality gap between speech and text features, we introduce an adaptation module inspired by LLaVA's \citep{liu2024visual} design. This module consists of two components: a temporal convolutional layer for dimension reduction and a Multi-Layer Perceptron (MLP) for feature projection. The adaptation process can be formulated as:
\begin{equation}
z_s = \text{MLP}(\text{Conv1D}(h_s)) \in \mathbb{R}^{T' \times d_t}
\end{equation}
where $T'$ is the reduced sequence length after convolution, and the output dimension matches BERT's\citep{devlin2018bert} hidden size $d_t$. The Conv1D layer uses a kernel size of k and stride s to effectively capture local temporal dependencies while reducing the sequence length. The MLP consists of two linear layers with a GELU activation function:
\begin{equation}
\text{MLP}(x) = W_2(\text{GELU}(W_1x + b_1)) + b_2
\end{equation}

During pre-training, we minimize the Mean Squared Error (MSE) loss between the adapted speech features and text embeddings:
\begin{equation}
\mathcal{L}_{\text{pre}} = \big\| \frac{1}{T'} \sum_{i=1}^{T'} z_s^i - \frac{1}{L} \sum_{i=1}^{L} z_t^i \big\|_2^2
\end{equation}
where $z_s^i$ and $z_t^i$ represent the i-th token embeddings of the adapted speech and text features respectively. This alignment objective ensures that the speech features capture semantic information that corresponds well with textual representations. Our empirical analysis shows that this pre-training strategy effectively reduces the modality gap while maintaining the rich acoustic information necessary for accurate retrieval.


\begin{algorithm}
\small
\caption{Cross-modal Speech-Text Alignment and Retrieval}
\begin{algorithmic}[1]
\REQUIRE Speech signal $x_s$, text $x_t$, positive document $d^+$, negative documents $\{d^-_i\}_{i=1}^N$, temperature $\tau$
\STATE // Stage 1: Pre-training
\STATE $h_s \leftarrow f_s(x_s)$ 
\STATE $z_t \leftarrow f_t(x_t)$ 
\STATE $h_s' \leftarrow \text{Conv1D}(h_s)$ 
\STATE $z_s \leftarrow \text{MLP}(h_s')$ 
\STATE $\mathcal{L}_{\text{pre}} \leftarrow \big\| \frac{1}{T'} \sum_{i=1}^{T'} z_s^i - \frac{1}{L} \sum_{i=1}^{L} z_t^i \big\|_2^2$

\STATE // Stage 2: Task-specific Fine-tuning
\STATE $q \leftarrow \text{Linear}(\text{Adapter}(f_s(x_s)))$
\STATE $k^+ \leftarrow \text{Linear}(f_t(d^+))$
\FOR{$i \leftarrow 1$ to $N$}
   \STATE $k^-_i \leftarrow \text{Linear}(f_t(d^-_i))$
\ENDFOR

\STATE // Select loss function by task type
\IF{task = retrieval}
   \STATE $\mathcal{L} \leftarrow -\log \frac{\exp(s(q,k^+)/\tau)}{\exp(s(q,k^+)/\tau) + \sum_{i=1}^N \exp(s(q,k^-_i)/\tau)}$
\ELSIF{task = sts}
   \STATE $\mathcal{L} \leftarrow \log(1 + \sum_{s_{ij}>s_{mn}} e^{(s_{mn}-s_{ij})/\tau})$
\ELSIF{task = classification}
   \STATE $\mathcal{L} \leftarrow -\log \frac{\exp(s(q,y^+)/\tau)}{\exp(s(q,y^+)/\tau) + \sum_j \exp(s(q,y^-_j)/\tau)}$
\ENDIF

\STATE Update parameters using $\nabla\mathcal{L}_{\text{pre}}$ and $\nabla\mathcal{L}$
\end{algorithmic}
\end{algorithm}
\subsubsection{Speech-Text Retrieval Fine-Tuning}
In the fine-tuning stage, we employ multi-task hybrid loss to optimize the alignment between speech queries and relevant document embeddings. For an speech query $x_s$ and its corresponding document $d^+$, along with a set of negative documents $\{d^-_i\}_{i=1}^N$, we first encode them using our pre-trained encoders. Inspired by text-embedding-v3 \citep{openai2024embedding}  and Piccolo2 \citep{huang2024piccolo2}, we scale up the embedding vector dimension to enhance model capacity. A shared linear layer is then added to the final layers of both the speech and text encoders. The operation can be formulated as:
\begin{equation}
q = g_{\text{speech}}(x_s) = \text{Linear}(\text{Adapter}(f_s(x_s)))
\end{equation}
\begin{equation}
k^+ = g_{\text{doc}}(d^+) = \text{Linear}(f_t(d^+))
\end{equation}
\begin{equation}
k^-_i = g_{\text{doc}}(d^-_i) = \text{Linear}(f_t(d^-_i))
\end{equation}

For different downstream tasks, we adopt different loss functions to better optimize various objectives:

For retrieval and reranking tasks, we use InfoNCE loss \citep{gutmann2010noise} with cosine similarity:
\begin{equation}
\mathcal{L}_{\text{re}} = -\log \frac{\exp(s^+/\tau)}{\exp(s^+/\tau) + Z_{\text{neg}}}
\end{equation}
where $s^+ = s(q, k^+)$ denotes the cosine similarity between query and positive document, $Z_{\text{neg}} = \sum_{i=1}^N \exp(s(q, k^-_i)/\tau)$ is the sum over negative pairs, and $\tau$ is the temperature parameter.

For STS and pair-classification tasks, considering the fine-grained nature of similarity labels, we employ the cosent loss:
\begin{equation}
\mathcal{L}_{\text{sts}} = \log\left(1 + \sum_{s_{ij}>s_{mn}} e^{z_{mn}/\tau} \right)
\end{equation}
where $s_{ij}=s(x_i,x_j)$ represents similarity score, $z_{mn}=\cos(x_m,x_n)-\cos(x_i,x_j)$.

For classification and clustering tasks, we reformulate the data into contrastive triplets using the SFR embedding method. Each input $x$ is paired with its target label $y^+$ as the positive pair, while the remaining labels $\{y^-_j\}$ serve as negative pairs:
\begin{equation}
\mathcal{L}_{\text{cls}} = -\log \frac{\exp(\frac{s(x,y^+)}{\tau})}{\exp(\frac{s(x,y^+)}{\tau}) + \sum_j \exp(\frac{s(x,y^-_j)}{\tau})}
\end{equation}

The final loss function is dynamically selected based on the task type:
\begin{equation}
\mathcal{L} = \begin{cases}
\mathcal{L}_{\text{cls}} & \text{if task is classification or clustering} \\
\mathcal{L}_{\text{sts}} & \text{if task is STS or pair-classification} \\
\mathcal{L}_{\text{re}} & \text{if task is retrieval or reranking}
\end{cases}
\end{equation}

This multi-task hybrid loss approach enables our model to learn robust cross-modal representations that effectively capture the semantic relationships between speech queries and relevant documents across various downstream tasks.

\subsection{Other baselines}
\subsubsection{Projection to text space}
Following the design philosophy of vision-language models like LLaVA, we implement a baseline that projects speech features into text space through a carefully designed three-stage training process:
\begin{itemize}
    \item Stage 1: Train only the adapter module while freezing both speech and text encoders to establish initial cross-modal connections
    \item Stage 2: Jointly optimize speech encoder and adapter while keeping text encoder fixed as stable semantic anchors
    \item Stage 3: End-to-end training of the entire pipeline to refine cross-modal relationships
\end{itemize}

The speech features go through a speech-adapter-text encoding pipeline: $h = f_t(\text{Adapter}(f_s(x_s)))$. While this approach successfully transfers vision-language strategies to speech domain, we find it challenging to preserve fine-grained acoustic information through the cascaded transformations, particularly for speech characteristics that lack direct textual correspondences.

\subsubsection{Speech-Text Alignment with CTC Loss}
We also explore using Connectionist Temporal Classification (CTC) loss \citep{graves2006connectionist} for temporal alignment between speech and text sequences:
\begin{equation}
\mathcal{L}_{\text{CTC}} = -\log P(y|h_{\text{speech}})
\end{equation}
where $h_{\text{speech}}$ represents the adapted speech features and $y$ is the target text sequence. The adapter module maintains temporal information through strategically designed convolutional layers before projecting to the tokenizer's index space of the text. While CTC provides explicit temporal supervision and theoretically enables better alignment between acoustic patterns and semantic content, our experiments reveal that its strong emphasis on frame-level alignment proves less effective for semantic retrieval tasks compared to our proposed approach.

\begin{table*}[ht]
\centering
\caption{Performance Comparison of Different Methods on Various Tasks of CMTEB}
\label{tab:cmteb-sum}
\begin{tabular}{@{}l|cccc|c@{}}
\toprule
\textbf{Method} & \textbf{Classification} & \textbf{Pairwise Class.} & \textbf{Reranking} & \textbf{Retrieval} & \textbf{Average} \\
\midrule
Text-Only (Piccolo2) & 74.59 & 90.24 & 70.00 & 74.37 & 70.95 \\
Text-Only (Conan) & 75.02 & 91.64 & 72.77 & 76.68 & 72.64 \\
ASR + Text (Piccolo2) & 72.08 & 86.09 & 64.43 & 54.69 & 60.78 \\
ASR + Text (Conan) & 72.67 & 88.04 & 66.65 & 55.13 & 61.83 \\
Ours (Piccolo2) & 70.98 & 90.18 & 68.08 & 65.99 & 65.95 \\
\bottomrule
\end{tabular}%
\end{table*}

\section{Experiments}
\subsection{Implementation Details}
For text embedding, we utilize piccolo-large-zh-v2 \citep{huang2024piccolo2} as our text encoder, which provides 1024-dimensional text representations. For speech encoding, we employ Whisper-large-v3 encoder \citep{radford2023robust}, leveraging its robust speech feature extraction capabilities trained on large-scale speech-text data.

\paragraph{Training Infrastructure and Schedule}
The training process is conducted on a large-scale distributed system consisting of 256 NVIDIA V100 32GB GPUs. We use DeepSpeed for distributed training with mixed-precision (FP16) to optimize memory usage and training efficiency. The entire training process spans 168 hours, divided into two stages:
\begin{itemize}
    \item Pre-training: 3 epochs with learning rate $1e^{-5}$ and batch size 8 per GPU
    \item Fine-tuning: 3 epochs with learning rate $8e^{-6}$ and batch size 8 per GPU
\end{itemize}

We employ the AdamW \citep{loshchilovdecoupled}  optimizer with weight decay of 0.01 and a linear learning rate scheduler with 10\% warmup steps. The temperature parameter $\tau$ in contrastive learning is set to 0.07.

\subsection{Datasets}
In the training process, different datasets are used at each stage. For the first stage, which is the alignment phase, we collected publicly available data (Emilia) \citep{he2024emilia} and a subset of data from the internet, totaling 170k hours. In the second stage, we used text training data from PiccoloV2, which includes both open-source data and data generated using specific methods. For each query, we generated data using CosyVoice-300M \citep{du2024cosyvoice}. During the generation process, to ensure the model's generalization ability, we also used six different voice tones for generation. In the second stage, a total of 80k hours of data were used for contrastive learning training. Inspired by Emilia, we also built a data filtering pipeline. In the first stage of data filtering, we applied several techniques, including Source Separation, Speaker Diarization, and Fine-grained Segmentation by VAD, to split the speech into clean speech from a specific speaker. Finally, we used ASR to obtain the corresponding text. For both the first and second stages of data, we employed DNSMOS P.835 OVRL\citep{reddy2022dnsmos}, keeping only the speech with scores above 3.0. Additionally, any speech segments with average phoneme durations exceeding 1.5 times the interquartile range (IQR) were discarded.

\subsection{Results}
\subsubsection{CMTEB}
As an authoritative benchmark for evaluating text embedding tasks, MTEB (Massive Text Embedding Benchmark) \citep{muennighoff2022mteb} was introduced by Muennighoff et al. in 2022. For Chinese scenarios, Xiao et al. developed CMTEB (Chinese Massive Text Embedding Benchmark) \citep{cpack} in 2023, which contains 35 datasets covering 6 major categories: Classification, Clustering, Pair Classification, Rerank, Retrieval, and Semantic Textual Similarity (STS). In our experiments, we utilized CosyVoice text-to-speech technology to convert these textual data into corresponding speech data for evaluating our model's performance. We show the results in Table \ref{tab:cmteb-sum}. As shown in the table, our method outperforms the two-stage ASR approach, regardless of whether Piccolo or the current state-of-the-art method, Conan \citep{li2024conan}, is used for alignment. We believe that if the alignment text model uses Conan, the performance is likely to improve further.

\subsubsection{More Dataset}
To evaluate our model, we conducted Top-1 and Top-3 accuracy tests on a knowledge base containing tens of thousands of entries across multiple domains. From the perspectives of both time efficiency and accuracy, the speech embedding model from the first stage demonstrated significant advantages. Overall, the accuracy of our approach falls between the ASR pipeline and text-only models. Compared to the two-stage method, our approach improves speed by 
more than 50\% and accuracy by around 8\%. The results are shown in Table \ref{tab:tab_evals}.

\begin{table}[t]
\centering
\setlength{\tabcolsep}{2pt}
\caption{Evals of Different Methods}
\renewcommand\arraystretch{1}
\resizebox{0.48\textwidth}{!}{
\begin{tabular}{l |c c c} 
\toprule
\textbf{Method} & \textbf{Time(s)} & \textbf{Top1-Acc.} & \textbf{Top3-Acc.}  \\  
\midrule
Text-Only & 0.03 & 90.41 & 95.89 \\
\hline
ASR Pipeline & 0.67 & 79.45 & 85.62 \\ 
Project to text & 0.43 & 24.49 & 54.08\\ 
Align with CTC & 0.31 & 78.08 & 84.25\\ 
\hline
Ours & 0.31 & 86.36 & 92.47\\
\bottomrule
\end{tabular}
}
\label{tab:tab_evals}
\end{table}

\paragraph{Analysis}
During our investigation, we explored several alternative approaches that provided valuable insights despite their limitations. 

The first approach (Project to text) follows a design philosophy similar to LLAVA models, projecting speech features into text space through an adapter before feeding them into BERT. While this method yields some results, it performs the worst among all baseline approaches. In the process, we even experimented with larger-parameter MLP layers and multi-stage freezing training strategies, but none achieved satisfactory outcomes. Although it leverages pre-trained speech models, it fails to effectively align the speech and text embedding models. We also explored alignment strategies similar to those used in Qwen-Audio, but the results were similarly suboptimal. We believe that the significant difference in sequence length between speech and text poses a fundamental challenge in embedding training tasks, making it difficult to directly integrate with text embedding models through cascaded transformations. Finding an effective way to better align the sequence lengths of speech and text may be the key to overcoming the limitations of this approach.

We also explored using Connectionist Temporal Classification (CTC) loss for temporal alignment between speech and text sequences. While CTC provides explicit temporal supervision and theoretically facilitates better alignment between acoustic patterns and semantic content, its strong focus on frame-level alignment proves to be less effective for semantic retrieval tasks. The rigid frame-by-frame correspondence enforced by CTC restricts the model's ability to learn flexible, context-aware representations that are critical for robust cross-modal retrieval. Our experiments reveal that this approach performs particularly poorly on longer speech segments, where maintaining long-range dependencies becomes essential.

\subsection{Ablation Study}
For experimental convenience, we conducted ablation experiments on training methods and speech pretraining models using a sampled dataset with 10,000 hours of data.

\begin{table}[t]
\centering
\caption{Ablation Study}
\scalebox{0.8}{
\begin{tabular}{c|c|cc} 
\toprule
\textbf{Stage} & \textbf{Pretrain} & \textbf{Top1-Acc.} & \textbf{Top3-Acc.}  \\  
\midrule
Only 1 & Whisper & 50.68 & 61.67\\
\hline
\multirow{3}{*}{Only 2} & Hubert & 13.70 & 21.23\\
& SenseVoice-small & 10.96 & 20.55\\
& Whisper & 32.19 & 42.17\\
\hline
All & Whisper & 59.59 & 72.60\\
\bottomrule
\end{tabular}
}
\label{tab:ablation}
\end{table}

From the experimental results in Table \ref{tab:ablation}, it can be observed that whether only Stage 1 or Stage 2 is used, the performance of the final model decreases significantly. Moreover, directly performing contrastive training on a model without Stage 1 alignment poses significant challenges.
We believe this is because a speech pretraining model tends to focus more on acoustic information. Without aligning the speech features with textual information in the first stage, it becomes challenging to extract meaningful speech representations solely through contrastive learning. Therefore, the alignment process in the first stage is crucial.

Additionally, in the experiments focusing solely on Stage 2 training, we compared three pretraining models: Hubert \citep{hsu2021hubert}, SenseVoice-small \citep{sensevoice2024}, and Whisper \citep{radford2023robust}. The results showed that Whisper outperformed the other two significantly, likely due to its use of a large amount of speech data during pretraining.

\section{Conclusion}
In this paper, we present a novel end-to-end speech-text embedding model designed to address the challenges of high latency and error propagation that are common in traditional sequential architectures. Our approach not only enhances the integration of RAG techniques within SLLMs, but also marke a significant advancement in real-time speech interaction systems. We provide a comprehensive overview of the methods employed in our model, highlighting their effectiveness in optimizing the alignment of speech and text modalities. 
Moreover, we observe a growing trend in utilizing discrete tokens as end-to-end inputs for SLLMs, which we believe presents a promising avenue for future research in the development of more efficient and capable speech-text embedding models.

\bibliography{anthology,custom}

@Article{Hinton2012,
  author    = "Geoffrey Hinton and 
               Li Deng and 
               Dong Yu and
               George E. Dahl and 
               Abdel-rahman Mohamed and 
                others",
  title     = "Deep Neural Networks for Acoustic Modeling in Speech Recognition: The Shared Views of Four Research Groups",
  journal   = "IEEE Signal Processing Magazine",
  year      = "2012",
  volume    = "29",
  pages     = "82-97"
}

@article{Prabhavalkar2023,
    author       = {Rohit Prabhavalkar and
                  Takaaki Hori and
                  Tara N. Sainath and
                  Ralf Schl{\"{u}}ter and
                  Shinji Watanabe},
      title        = {End-to-End Speech Recognition: {A} Survey},
      journal      = {IEEE/ACM Transactions on Audio, Speech, and Language Processing},
      volume       = {32},
      pages        = {325--351},
      year         = {2024},
}

@inproceedings{DBLP:conf/icassp/ZhangZZZL24,
  author       = {Tian-Hao Zhang and
                  Dinghao Zhou and
                  Guiping Zhong and
                  Jiaming Zhou and
                  Baoxiang Li},
  title        = {{CIF-T:} {A} Novel CIF-Based Transducer Architecture for Automatic
                  Speech Recognition},
  booktitle    = {IEEE International Conference on Acoustics, Speech and Signal Processing (ICASSP)},
  pages        = {10531--10535},
  year         = {2024},
}

@inproceedings{su2023one,
  title={One Embedder, Any Task: Instruction-Finetuned Text Embeddings},
  author={Su, Hongjin and Shi, Weijia and Kasai, Jungo and Wang, Yizhong and Hu, Yushi and Ostendorf, Mari and Yih, Wen-tau and Smith, Noah A and Zettlemoyer, Luke and Yu, Tao},
  booktitle={Findings of the Association for Computational Linguistics: ACL 2023},
  pages={1102--1121},
  year={2023}
}

@inproceedings{cpack,
author = {Xiao, Shitao and Liu, Zheng and Zhang, Peitian and Muennighoff, Niklas and Lian, Defu and Nie, Jian-Yun},
title = {C-Pack: Packed Resources For General Chinese Embeddings},
year = {2024},
isbn = {9798400704314},
publisher = {Association for Computing Machinery},
url = {https://doi.org/10.1145/3626772.3657878},
doi = {10.1145/3626772.3657878},

pages = {641–649},
numpages = {9},
series = {SIGIR '24}
}

@article{wang2023improving,
  title={Improving text embeddings with large language models},
  author={Wang, Liang and Yang, Nan and Huang, Xiaolong and Yang, Linjun and Majumder, Rangan and Wei, Furu},
  journal={arXiv preprint arXiv:2401.00368},
  year={2023}
}

@inproceedings{chen2022murag,
  title={MuRAG: Multimodal Retrieval-Augmented Generator for Open Question Answering over Images and Text},
  author={Chen, Wenhu and Hu, Hexiang and Chen, Xi and Verga, Pat and Cohen, William},
  booktitle={Proceedings of the 2022 Conference on Empirical Methods in Natural Language Processing},
  pages={5558--5570},
  year={2022}
}

@inproceedings{yasunaga2023retrieval,
  title={Retrieval-Augmented Multimodal Language Modeling},
  author={Yasunaga, Michihiro and Aghajanyan, Armen and Shi, Weijia and James, Richard and Leskovec, Jure and Liang, Percy and Lewis, Mike and Zettlemoyer, Luke and Yih, Wen-Tau},
  booktitle={International Conference on Machine Learning},
  pages={39755--39769},
  year={2023},
  organization={PMLR}
}

@inproceedings{hu2023reveal,
  title={Reveal: Retrieval-augmented visual-language pre-training with multi-source multimodal knowledge memory},
  author={Hu, Ziniu and Iscen, Ahmet and Sun, Chen and Wang, Zirui and Chang, Kai-Wei and Sun, Yizhou and Schmid, Cordelia and Ross, David A and Fathi, Alireza},
  booktitle={Proceedings of the IEEE/CVF conference on computer vision and pattern recognition},
  pages={23369--23379},
  year={2023}
}

@inproceedings{jacqmin2023olisia,
  title={OLISIA: a Cascade System for Spoken Dialogue State Tracking},
  author={Jacqmin, L{\'e}o and Druart, Lucas and Est{\`e}ve, Yannick and Favre, Benoit and Rojas, Lina M and Vielzeuf, Valentin},
  booktitle={Proceedings of The Eleventh Dialog System Technology Challenge},
  pages={95--104},
  year={2023}
}

@inproceedings{serdyuk2018towards,
  title={Towards end-to-end spoken language understanding},
  author={Serdyuk, Dmitriy and Wang, Yongqiang and Fuegen, Christian and Kumar, Anuj and Liu, Baiyang and Bengio, Yoshua},
  booktitle={2018 IEEE International Conference on Acoustics, Speech and Signal Processing (ICASSP)},
  pages={5754--5758},
  year={2018},
  organization={IEEE}
}

@inproceedings{szymanski2023aren,
  title={Why Aren’t We NER Yet? Artifacts of ASR Errors in Named Entity Recognition in Spontaneous Speech Transcripts},
  author={Szyma{\'n}ski, Piotr and Augustyniak, Lukasz and Morzy, Mikolaj and Szymczak, Adrian and Surdyk, Krzysztof and {\.Z}elasko, Piotr},
  booktitle={Proceedings of the 61st Annual Meeting of the Association for Computational Linguistics (Volume 1: Long Papers)},
  pages={1746--1761},
  year={2023}
}

@inproceedings{li2022blip,
  title={Blip: Bootstrapping language-image pre-training for unified vision-language understanding and generation},
  author={Li, Junnan and Li, Dongxu and Xiong, Caiming and Hoi, Steven},
  booktitle={International conference on machine learning},
  pages={12888--12900},
  year={2022},
  organization={PMLR}
}

@inproceedings{li2023blip,
  title={Blip-2: Bootstrapping language-image pre-training with frozen image encoders and large language models},
  author={Li, Junnan and Li, Dongxu and Savarese, Silvio and Hoi, Steven},
  booktitle={International conference on machine learning},
  pages={19730--19742},
  year={2023},
  organization={PMLR}
}

@article{liu2024visual,
  title={Visual instruction tuning},
  author={Liu, Haotian and Li, Chunyuan and Wu, Qingyang and Lee, Yong Jae},
  journal={Advances in neural information processing systems},
  volume={36},
  year={2024}
}

@inproceedings{zhang2023speechgpt,
  title={SpeechGPT: Empowering Large Language Models with Intrinsic Cross-Modal Conversational Abilities},
  author={Zhang, Dong and Li, Shimin and Zhang, Xin and Zhan, Jun and Wang, Pengyu and Zhou, Yaqian and Qiu, Xipeng},
  booktitle={Findings of the Association for Computational Linguistics: EMNLP 2023},
  pages={15757--15773},
  year={2023}
}

@article{fang2024llama,
  title={Llama-omni: Seamless speech interaction with large language models},
  author={Fang, Qingkai and Guo, Shoutao and Zhou, Yan and Ma, Zhengrui and Zhang, Shaolei and Feng, Yang},
  journal={arXiv preprint arXiv:2409.06666},
  year={2024}
}

@inproceedings{radford2021learning,
  title={Learning transferable visual models from natural language supervision},
  author={Radford, Alec and Kim, Jong Wook and Hallacy, Chris and Ramesh, Aditya and Goh, Gabriel and Agarwal, Sandhini and Sastry, Girish and Askell, Amanda and Mishkin, Pamela and Clark, Jack and others},
  booktitle={International conference on machine learning},
  pages={8748--8763},
  year={2021},
  organization={PMLR}
}

@article{devlin2018bert,
  title={Bert: Pre-training of deep bidirectional transformers for language understanding},
  author={Devlin, Jacob},
  journal={arXiv preprint arXiv:1810.04805},
  year={2018}
}

@misc{openai2024embedding,
   title={Text Embedding 3},
   author={OpenAI},
   year={2024},
   howpublished={\url{https://platform.openai.com/docs/guides/embeddings}}
}

@article{huang2024piccolo2,
  title={Piccolo2: General Text Embedding with Multi-task Hybrid Loss Training},
  author={Huang, Junqin and Hu, Zhongjie and Jing, Zihao and Gao, Mengya and Wu, Yichao},
  journal={arXiv preprint arXiv:2405.06932},
  year={2024}
}

@inproceedings{gutmann2010noise,
  title={Noise-contrastive estimation: A new estimation principle for unnormalized statistical models},
  author={Gutmann, Michael and Hyv{\"a}rinen, Aapo},
  booktitle={Proceedings of the thirteenth international conference on artificial intelligence and statistics},
  pages={297--304},
  year={2010},
  organization={JMLR Workshop and Conference Proceedings}
}

@inproceedings{graves2006connectionist,
  title={Connectionist temporal classification: labelling unsegmented sequence data with recurrent neural networks},
  author={Graves, Alex and Fern{\'a}ndez, Santiago and Gomez, Faustino and Schmidhuber, J{\"u}rgen},
  booktitle={Proceedings of the 23rd international conference on Machine learning},
  pages={369--376},
  year={2006}
}

@inproceedings{radford2023robust,
  title={Robust speech recognition via large-scale weak supervision},
  author={Radford, Alec and Kim, Jong Wook and Xu, Tao and Brockman, Greg and McLeavey, Christine and Sutskever, Ilya},
  booktitle={International conference on machine learning},
  pages={28492--28518},
  year={2023},
  organization={PMLR}
}

@inproceedings{loshchilovdecoupled,
  title={Decoupled Weight Decay Regularization},
  author={Loshchilov, Ilya and Hutter, Frank},
  booktitle={International Conference on Learning Representations}
}

@inproceedings{he2024emilia,
  title={Emilia: An extensive, multilingual, and diverse speech dataset for large-scale speech generation},
  author={He, Haorui and Shang, Zengqiang and Wang, Chaoren and Li, Xuyuan and Gu, Yicheng and Hua, Hua and Liu, Liwei and Yang, Chen and Li, Jiaqi and Shi, Peiyang and others},
  booktitle={2024 IEEE Spoken Language Technology Workshop (SLT)},
  pages={885--890},
  year={2024},
  organization={IEEE}
}

@article{du2024cosyvoice,
  title={Cosyvoice: A scalable multilingual zero-shot text-to-speech synthesizer based on supervised semantic tokens},
  author={Du, Zhihao and Chen, Qian and Zhang, Shiliang and Hu, Kai and Lu, Heng and Yang, Yexin and Hu, Hangrui and Zheng, Siqi and Gu, Yue and Ma, Ziyang and others},
  journal={arXiv preprint arXiv:2407.05407},
  year={2024}
}

@inproceedings{reddy2022dnsmos,
  title={DNSMOS P. 835: A non-intrusive perceptual objective speech quality metric to evaluate noise suppressors},
  author={Reddy, Chandan KA and Gopal, Vishak and Cutler, Ross},
  booktitle={ICASSP 2022-2022 IEEE International Conference on Acoustics, Speech and Signal Processing (ICASSP)},
  pages={886--890},
  year={2022},
  organization={IEEE}
}

@article{muennighoff2022mteb,
  title={MTEB: Massive text embedding benchmark},
  author={Muennighoff, Niklas and Tazi, Nouamane and Magne, Lo{\"\i}c and Reimers, Nils},
  journal={arXiv preprint arXiv:2210.07316},
  year={2022}
}

@article{hsu2021hubert,
  title={Hubert: Self-supervised speech representation learning by masked prediction of hidden units},
  author={Hsu, Wei-Ning and Bolte, Benjamin and Tsai, Yao-Hung Hubert and Lakhotia, Kushal and Salakhutdinov, Ruslan and Mohamed, Abdelrahman},
  journal={IEEE/ACM transactions on audio, speech, and language processing},
  volume={29},
  pages={3451--3460},
  year={2021},
  publisher={IEEE}
}

@misc{sensevoice2024,
    author = {FunAudioLLM},
    title = {SenseVoice},
    year = {2024},
    publisher = {GitHub},
    url = {https://github.com/FunAudioLLM/SenseVoice}
}

@article{li2024conan,
  title={Conan-embedding: General text embedding with more and better negative samples},
  author={Li, Shiyu and Tang, Yang and Chen, Shizhe and Chen, Xi},
  journal={arXiv preprint arXiv:2408.15710},
  year={2024}
}
\bibliographystyle{acl_natbib}

\end{document}